\documentclass[prl,twocolumn,showpacs,amsmath,amssymb,superscriptaddress,floatfix]{revtex4-1}

\usepackage[english]{babel}
\usepackage{amsfonts}
\usepackage{graphicx}
\usepackage{color}

\newcommand{\fmf}{\affiliation{Faculty of Mathematics and Physics, University of Ljubljana, SI-1000 Ljubljana, Slovenia}}
\newcommand{\ijs}{\affiliation{J. Stefan Institute, SI-1000 Ljubljana, Slovenia}}
\newcommand{\lmu}{\affiliation{Department of Physics and Arnold Sommerfeld Center for Theoretical Physics, Ludwig-Maximilians-Universit\"at M\"unchen, D-80333 M\"unchen, Germany}}
\newcommand{\katowice}{\affiliation{Department of Theoretical Physics, Institute of Physics, University of Silesia, 40Ð007 Katowice, Poland}}

\begin{document}

\title{Nonequilibrium propagation and decay of a bound pair in driven $t$--$J$ models}

\author{J. \surname{Bon\v ca}}
\fmf
\ijs

\author{M. \surname{Mierzejewski}}
\katowice

\author{L. \surname{Vidmar}}
 \email[E-mail: ]{Lev.Vidmar@lmu.de}
\ijs
\lmu

%
%

\date{\today}
\begin{abstract}
We perform an accurate time--dependent numerical study of out--of--equilibrium response of a bound state within $t$--$J$ systems on a two--leg ladder and a square lattice.
We show that the bound hole pair decays with the onset of finite steady current if both mechanisms for binding and the dissipation share matching degrees of freedom.
Moreover, by investigating the mechanism of decay on the square lattice we find that the dynamics is governed by the decay in the direction perpendicular to the electric field, leading to much shorter decay times in comparison to the ladder where such dynamics is topologically restricted.
\end{abstract}

\maketitle

\textit{Introduction.}---%
Equilibrium states of correlated many--body systems can exhibit many distinct collective phenomena, with the Mott insulating, superconducting (SC) and magnetically ordered states being the most prominent examples.
A fundamental question in this respect is how such many--body states behave when driven far from their equilibrium by finite external fields.
Many theoretical studies  have recently focused on the dielectric breakdown of the Mott insulator~\cite{oka05,*heidrichmeisner10,*kirino10,*zala12,*aron12,eckstein10a} and the majority of them addressed this problem within the framework of the half--filled Hubbard model.
Much less is known about nonequilibrium properties of strongly--correlated systems away from half--filling, where unconventional superconductivity may possibly emerge.
Ballistic response of superconductors to constant electric field has recently been observed~\cite{saracila09}.
It shows up as a current that increases linearly in time up to a threshold value above which the superconductivity is destroyed.
The properties of cuprate superconductors subject to the electric current have also been studied within a variational (equilibrium) 
approach~\cite{goren10}.
For low concentration of holes, sufficiently strong current destroys superfluid stiffness while pairing remains intact. 

In this Letter, we do not discuss superconductivity.
Instead we focus on the real--time nonequilibrium dynamics of its basic ingredients, i.e., bound pairs of charge carriers.  
According to the Ohm's law, driving of charge carriers by constant electric field leads to a finite time--independent current due to a steady emission of excitations, e.g., phonons or magnons.
Here, we address a question whether a bound pair of two carriers can respond  in a similar way, i.e., whether it can acquire a constant velocity upon constant electric field without decaying into two separate carriers.
In principle, such problem can be addressed in experiments on systems in which pairing precedes the SC phase coherence.

Despite a seeming simplicity of this problem the answer is not immediately obvious since pairing degrees of freedom simultaneously present quite effective dissipation channels.
Therefore, propagation of a bound pair under electric field causes steady  emission (heating) of magnons or spin fluctuations which simultaneously mediate the pairing interaction.
Here, we consider two holes in  the $t$--$J$ ladder and square lattice and carry out fully microscopic calculations taking into account two  most relevant phenomena:
{\it (i)} pairing by the exchange of spin excitations that has been so far studied predominantly under equilibrium conditions~\cite{*[{See, e.g., }] [{ and references therein; }] dagotto_rev,*chernyshev98,*wrobel98,*riera98,*barentzen99,*leung02,*lau11b, [{See also }] wellein96,*sakai97,*hague07,*huang11,*alexandrov12,vidmar09b,*maska12};
{\it (ii)} dissipation by emission of spin excitations~\cite{vidmar11b}.
Recent investigations of driven systems at half--filling have contributed to a general understanding of heating in isolated systems which on a long time scale suppresses any steady current~\cite{marcin10,eckstein11b}.
There are two possible ways to avoid this problem:
either to couple the system to enivironment~\cite{fukui98,*sugimoto08,amaricci11, aron11,balzer11,*knap11, alvermann12}, or to consider
vanishingly small concentration of charge carriers~\cite{vidmar11b,vidmar11c}.
In this Letter, we consider the second option.
For a fixed number of carriers (two holes) heating is a finite size effect~\cite{vidmar11b}; thus we restrict our analysis to regimes where the results are essentially size independent.

Another important property of driven quantum systems that remains a widely unexplored subject is the role of dimensionality transpiring in the compelling phenomena emerging in the direction perpendicular to the driving~\cite{aron11,andre12}.
The most common numerical approaches to nonequilibrium correlated systems were developed for studies of either one--dimensional~\cite{feiguin04,*daley04} or high--dimensional~\cite{monien02,*freericks06,schiro10} systems.
By using dynamical mean--field theory it has been shown that in the limit of extremely strong electric field, 
a $D$--dimensional system exhibits equilibrium properties in $D-1$ dimensions perpendicular to the field~\cite{aron11}.
We show that for moderate electric field an unexpected effect emerges in the 2D system: the decay of the bound state is governed by the motion of charge carriers perpendicular to the field.

\textit{Model and setup.}---%
We consider a driven $t$--$J$ model with two holes (also referred to as charge carriers) on the ladder and the square lattice
\begin{equation}
H = -t_0 \sum_{\langle {\bf ij}\rangle,s}(\tilde c^\dagger_{{\bf i},s} \tilde c_{{\bf j},s} e^{i\phi_{\bf ij}(t)} +\mathrm{H.c.}) +
\sum_{\langle {\bf ij}\rangle } J_{\bf ij} ( {\bf S}_{\bf i} {\bf S}_{\bf j} - \frac{1}{4}\tilde{n}_{\bf i} \tilde{n}_{\bf j} )
\label{ham}
\end{equation}
where $\tilde c_{{\bf i},s} = c_{{\bf i},s}(1 - n_{{\bf i},-s})$ is a projected fermion operator, $t_0$ represents nearest neighbor overlap integral, the sum $\langle \bf ij \rangle$  runs over pairs of nearest neighbors and
$\tilde n_{\bf i} = n_{\bf{i},\uparrow} + n_{\bf{i},\downarrow} - 2 n_{\bf{i},\uparrow} n_{\bf{i},\downarrow}$ is a projected electron number operator.
On the ladder, $J_{\bf ij}$ may be different for interactions along and perpendicular to the ladder's leg, while we set $J_{\bf ij}=J$ for the square lattice.
The constant electric field $F$ is switched on at $t=0$.
It is applied along the ladder's leg and along the diagonal of the square lattice, i.e.,
we set $\phi_{\bf ij}(t)=-(+)Ft$ and $\phi_{\bf ij}(t)=-(+)Ft/\sqrt{2}$ for positive (negative) directions of carrier hopping in Eq.~(\ref{ham}), respectively%
~\footnote{%
We measure $F$ in units of $[t_0/e_0 a]$ where $e_0$ is the unit charge and $a$ the lattice distance.
We measure time in units of $1/t_0$ and set $t_0=e_0=a=1$ throughout the work.
}.

We apply the time--dependent exact diagonalization method ($t$--ED) within the full Hilbert space to calculate out--of--equilibrium response of the driven $t$--$J$ ladder with periodic boundary conditions, while we use the time--dependent exact diagonalization method defined over a limited functional space ($t$--EDLFS) for the $t$--$J$ square lattice~\cite{bonca2,vidmar11b}.
The latter method has been successfully applied to calculation of the ground state of the $t$--$J$ model with two doped holes~\cite{vidmar09b,*maska12};
the details of the method are given elsewhere%
~\footnote{%
The construction of the functional space starts from a N\' eel state with two holes located on neighboring ${\rm Cu}$ sites~\cite{vidmar09b}, which represents a parent state of a translationally invariant state with $d$--wave symmetry and ${\bf k} = (0, 0)$.
We generate new parent states by applying the generator of states as described in Eq.~(3) of Ref.~\cite{bonca2} with $N_h = 14$.
Full Hamiltonian at time $t=0$ is diagonalized within the limited functional space taking explicitly into account translational symmetry.
}.

Time evolution of both systems subjected to an external electric field is calculated using the iterative Lanczos method~\cite{park86}.
This method has been lately applied to calculate out--of equilibrium response of different quantum many--body systems, both during the constant driving~\cite{marcin10,vidmar11c} as well as during and after photoexcitations~\cite{matsueda11,*defilippis12}.
In contrast to the large part of recent nonequilibrium studies of driven strongly--correlated systems~\cite{marcin10,freericks06,freericks08,*karlsson11,eckstein10a,eckstein11b}, our approach enables calculation of the steady state where the driven charge carriers acquire constant velocity due to the propagation in dissipative medium.
The latter, which may also be referred to as a quantum heat bath,
%
can be either modeled by interaction with magnons~\cite{vidmar11b}, phonons~\cite{vidmar11a,*golez11}, or both~\cite{vidmar11c}.
Interestingly, properties of these systems share some similarities with driven systems at half--filling, with the most prominent example being the current--field ($\bar\jmath$--$F$) characteristics~\cite{eckstein10a,amaricci11,aron11,vidmar11b,vidmar11c,einhellinger12}.
However, the calculation of $\bar\jmath$--$F$ characteristics is not the main goal of the present study.
Instead, we rather focus on the conditions and mechanism of decay of the bound state under the influence of the electric field.

\textit{Propagation of a bound state on the ladder.}---%
The response of a quantum system to a constant electric field considerably  depends on the strength of the field.
For very small $F\rightarrow 0$ the adiabatic regime (AR) with zero net current is observed, while larger $F$ gives rise to the dissipative regime (DR) where constant $F$ induces a finite
dc current~\cite{vidmar11a,vidmar11b}.
Throughout the work, we introduce the average distance between holes in parallel ($d_{\Vert}$) and perpendicular ($d_{\perp}$) direction, defined as
$$
d_{\mu} = \sqrt{ \sum_{\bf r} \left( {\bf r \cdot \hat{e}_{\mu}} \right)^2 {\cal C}({\bf r}) } - d_{\mu}(t=0),
$$
where
$
{\cal C}({\bf r}) = \sum_{\bf i}  \langle n_{\bf i}^h n_{\bf i+r}^h \rangle /2
$
($n_{\bf i}^h$ is a hole number operator),
while ${\bf\hat{e}}_{\Vert}$ and ${\bf\hat{e}}_{\perp}$ are unit vectors parallel and perpendicular to the field, respectively.

\begin{figure}[!htb]
\includegraphics[width=0.99\columnwidth,clip]{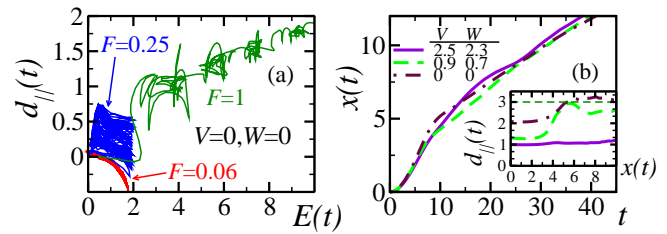}
\caption{(Color online)
Time evolution on a $2 \times 10$ ladder with two holes under constant $F$ switched on at $t=0$.
(a) $d_{\Vert}(t)$ vs $E(t)$ for $J_{\perp}=1$ (along the rungs), $J_{\parallel}=0.4$ (along the legs) and various $F$.
(b) $x(t)$ (main) and $d_{\Vert}(t)$ vs $x(t)$ (inset) for three different values of $V$ and $W$.
We use $J_{\perp}=J_{\parallel}=0.4$ and $F=0.125$ in (b).
In the inset, $d_{\Vert}(t)$ is not subtracted by $d_{\Vert}(t=0)$;
thin dashed horizontal line shows $d_{\parallel}$ for two noninteracting fermions.
}
\label{ladder}
\end{figure}

The existence of the bound pairs in AR is rather unambiguous.
If the spin gap (between  the ground state and the excited states) is finite for arbitrary momentum, then the system remains  in its instantaneous eigenstate.
The distance between carriers $d_{\parallel}(t)$ and the change of the total energy $E(t)$ are determined by the instantaneous value of the adiabatic parameter $\phi(t)=Ft$.
Since the Hamiltonian $H[\phi]$ is periodic,  $d_{\parallel}(t)$ and $E(t)$ are periodic as well. 
They oscillate with a frequency twice larger than the Bloch frequency $\omega_B=F$, where doubling originates from the (double) charge of the bound pair.
Plotting the system evolution in the plane $[E(t),d_{\parallel}(t)]$, closed loops emerge as a result of periodicity.
Such a loop is shown in Fig.~\ref{ladder}(a), see the (red) curve for the weakest field.
The horizontal extent  of this loop is determined by the dispersion of the ground state. Here, we show results for anisotropic $J$ when exchange interaction along the rungs $J_{\perp}$ is much stronger than the interaction along the legs $J_{\parallel}$.
In such case the spin gap occurs for arbitrary flux and the presence of AR for $F\rightarrow 0$ is unquestionable. 

Contrary to AR, DR is characterized by a steady increase of energy $E(t)$. Therefore, moving from AR to DR  must be accompanied by a destruction of the loops in the  $[E(t),d_{\parallel}(t)]$ plane.
The direction of the loop deformation shows whether bound pairs survive in DR. 
If they do, then the loops should be elongated only in the horizontal direction [$E(t)$ increases], while in the vertical direction $d_{\parallel}(t)$ should remain bound.
However, as shown in  Fig.~\ref{ladder}(a) the opposite happens: the increase of $d_{\parallel}(t)$ goes along with
the increase of energy.
This indicates that the bound pairs dissociate immediately when DR sets in.
Carrying out calculations for various parameters and fields, we found no case with propagating bound pairs in DR. 

The central question is why is it so and what mechanism would allow pairs to propagate with a steady velocity in a dissipative environment.  
The propagation with a steady velocity under a constant $F$ leads to a steady increase of energy. Since the average kinetic energy of charge carriers should remain constant (steady current in DR), the electrostatic energy has to be transformed into excitations of the spin background.
However, the interaction of holes with the spin background is simultaneously the only pairing mechanism. 

Below we demonstrate that this double role of spin excitations - as a dissipation mechanism and as a pairing mechanism - is responsible for decay of bound pairs in DR. For this sake we extend the $t$--$J$ model by the nearest ($V$) and next nearest neighbor ($W$) attractive interactions
\begin{equation}
H \rightarrow H - V \sum_{\langle {\bf ij} \rangle} n_{\bf i}^h n_{\bf j}^h  - W  \sum_{\langle \langle {\bf ij} \rangle \rangle} n_{\bf i}^h n_{\bf j}^h ,
\label{ext1}
\end{equation}
which play the role of additional pairing mechanisms.
Fig.~\ref{ladder}(b) demonstrates how this mechanism affects $d_{\parallel}(t)$ and the distance travelled by the center of mass $x(t)$ for various $V$ but constant $V-W$.
We notice that changing of $V$ and $W$ does not influence $x(t)$ and the steady increase of energy $E(t)=2 F x(t)$ being a hallmark of DR is clearly visible.
While in the pure $t$--$J$ model the pairs dissociate very quickly, see the curve $V=W=0$ in the inset of Fig.~\ref{ladder}(b), both holes stay
together for sufficiently large $V$ and $W$.
Therefore, we notice that the bound pair of carriers can propagate under constant $F$ with a steady velocity provided that there are different mechanisms (degrees of freedom) responsible for pairing and dissipation.
Introducing simultaneously $V$ and $W$ leads to the attractive potential between holes that allows hopping of the hole pair without breaking the attractive potential.
The opposite case of large $V$ and $W = 0$ leads to pair breaking.

Inclusion of the interactions $V$ and $W$ allowed us to explain why bound pairs in the pure $t$--$J$ model (an possibly also in the electron--phonon systems~\cite{golez11}) decay as soon  as the response is dominated by the dc current.
We would like to emphasize though that replacing attractive $V$ and $W$ terms with more realistic electron--phonon interaction would introduce additional channel for the dissipation of potential energy through emission of phonons.
It is thus plausible to speculate, that a spin-lattice bound pair would as well decay in the dissipative regime.
From now on, we again focus on the isotropic $t$--$J$ model and set $V=W=0$.

\begin{figure}[!htb]
\includegraphics[width=0.99\columnwidth,clip]{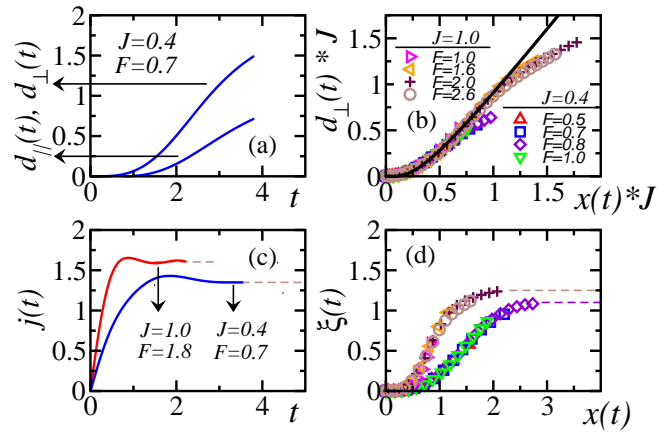}
\caption{(Color online)
Driven hole pair on the square lattice.
(a) $d_{\Vert}(t)$ and $d_{\perp}(t)$ for $J=0.4$ and $F=0.7$.
(b) $d_{\perp}(t)J$ vs $x(t)J$ for different $J$ and $F$.
Black solid line represents a fit $d_{\perp}(t) = \beta x(t) \exp{(-\frac{1}{\alpha J x(t)})}$, where $\beta = 1.474$ and $\alpha=2.066$.
(c) Current $j(t)$ for different $J$ and $F$.
(d) $\xi(t)$ vs $x(t)$, see Eq.~(\ref{ksi1}), for the same set of parameters as in (b).
}\label{fig2}
\end{figure}

\textit{Mechanism of decay.}---%
We now extend out study to the square lattice. We compare the response of the system in the parallel vs perpendicular direction with respect to  the electric field. 
When investigating properties of the bound state, see Fig.~\ref{fig2}(a), we observe $d_{\perp}(t) > d_{\Vert}(t)$ for all times of our calculation.
This implies, in contrast to the ladder system, that the dynamics perpendicular to the electric field governs the decay of the bound state for (at least) short and intermediate times.
We focus on this issue further on to clarify the mechanism of decay on the square lattice.

We first investigate the transient time $t<t^*$, where $t^*$ is defined as the characteristic time needed to reach the steady state after turning on the field.
We may expect that the dynamics of decay at $t<t^*$ is strongly dependent on properties of the bound state in equilibrium, which is determined by the energy scale $J$~\cite{vidmar09b}.
Remarkably, numerical data reveal both $J$-- and $F$--independent scaling of $d_{\perp}(t)$ vs $x(t)$ at short times.
As shown in Fig.~\ref{fig2}(b), the universal relation between $d_{\perp}(t)$ and $x(t)$ can be well described by
\begin{equation}
d_{\perp}(t) \propto x(t) \; e^{-\frac{1}{\alpha J x(t)}}, \label{activated}
\end{equation}
with $\alpha \sim 2$.
Such relation, which indicates an activated--type of behavior, incorporates the information about the size of the bound state at $t=0$.
Indeed, the average distance between two holes in the equilibrium scales roughly with $1/J$ for $J>0.4$%
~\footnote{%
For the $d$--wave bound state using the EDLFS method, the average distance $D$ between two holes in the $t$--$J$ model for $0.4<J<1.5$ scales approximately as $D = 1 + a J^{-b}$, where $a=0.33$ and $b=0.93$%
}.
For $t>t^*$
the scaling of Eq.~(\ref{activated}) breaks down,
however, $d_{\perp}(t)$ vs $x(t)$ still show a $F$--independent behavior for a fixed $J$.
In the following, we will show that such $F$--independent scaling is as well expected for larger times $t \gg t^*$.

\begin{figure*}[!htb]
\includegraphics[width=2\columnwidth,clip]{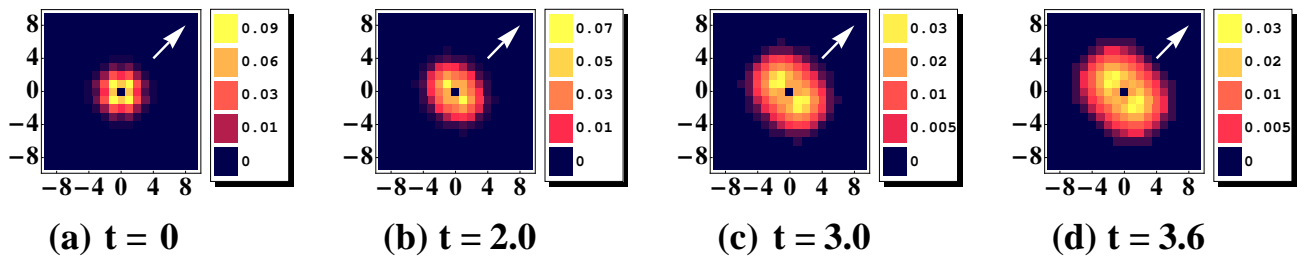}
\caption{(Color online) %
${\cal C}({\bf r})$ measuring time--dependent probability for the hole pair to be at a relative position ${\bf r}$, fulfilling the sum rule $\sum_{\bf r} {\cal C}({\bf r}) = 1$.
Arrows mark direction of electric field.
We set $J=0.4$ and $F=0.7$.
}\label{fig4}
\end{figure*}

We now proceed to describe the properties of the steady state, which is defined as the regime when the current along the field $j(t)$ is time independent.
In Fig.~\ref{fig2}(c) we plot $j(t)$ which clearly marks the onset of constant current for $t \gtrsim 2$.
In this regime holes are already well separated in space and the binding mechanism should be significantly weaker than in the equilibrium.
Therefore, one may expect a rather independent movement of holes  in the direction perpendicular to $F$. 
This movement can be naively modeled by a random--walking process: 
\begin{equation}
d^2_{\perp}(t) = d^2_{\perp}(t^*) + (t - t^*) D_{\perp}, \label{d2qs}
\end{equation}
where  $(t - t^*) D_{\perp}$ is proportional to the number of random steps in the time window $(t-t^*)$.
In the same way, we may define $x(t)$ for $t>t^*$ as
%
\begin{equation}
x(t) = x^* + (t-t^*) \; \bar\jmath, \label{xqs}
\end{equation}
where again $x^* = x(t^*)$ and $\bar\jmath$ represents the steady (dc) current.
Expressing $(t - t^*)$ from Eq.~(\ref{xqs}) and inserting it into Eq.~(\ref{d2qs}), 
we may express time--dependence of $d_{\perp}(t)$ through $x(t)$.
%
%
To justify the choice of ansatz in Eq.~(\ref{d2qs}), we introduce a quantity $\xi(t)$ defined as
\begin{equation}
\xi(t) = \frac{d^2_{\perp}(t)}{x(t)} = \frac{A}{x(t)} + \frac{D_{\perp}}{\bar\jmath}, \label{ksi1}
\end{equation}
where $A=d^2_{\perp}(t^*) - \frac{D_{\perp}}{\bar\jmath} x^*$ (in our case, $A<0$).
For long enough times, we expect $A/x(t) \to 0$ and $\xi(t)$ should approach the constant $\bar\xi = D_{\perp}/\bar\jmath$.
Indeed, we observe in Fig.~\ref{fig2}(d) the saturation of $\xi(t)$ for different values of $F$ and $J$.
Moreover, for a fixed $J$ the values of $\bar\xi$ are independent of the strength of electric field $F$.
This result implies that for moderate driving the number of random steps $(t - t^*) D_{\perp}$ is proportional to the distance travelled along the field $(t - t^*) \bar\jmath$.
This proportionality suggests yet another strong argument supporting the decay of the driven bound state in 2D.

Decay of the bound state can be also monitored by calculating the 2D correlation function
${\cal C}({\bf r})$ measuring time--dependent probability for the hole pair to be at a relative position ${\bf r}$.
Results in Fig.~\ref{fig4} show a disk--shaped pattern of ${\cal C}({\bf r})$ elongated perpendicular to the field, consistent with Fig.~\ref{fig2}(a).
Moreover, a perpendicular cut through ${\bf r}=(0,0)$ (not shown) reveals that the position of maximum of ${\cal C}({\bf r})$ steadily moves to larger $\vert {\bf r} \vert$,
determining the main direction of decay.

\textit{Discussion and Conclusion.}---%
By applying the $t$--ED and $t$--EDLFS method to study real--time response of a fully quantum system, we managed to follow the out of equilibrium dynamics of a driven system where initially at $t=0$ the bound state exists  due to the exchange of spin excitations. 
Our calculations on the ladder system show that as long as there is no additional mechanism to provide the glue for binding, the bound state of two charge carriers decays with the onset of finite steady current.
In the 2D system a bound pair decays predominantly  in the perpendicular direction with respect to the external field, which consequently allows for more efficient release of the gained potential energy through magnon emission.
At longer times, however, the motion of carriers perpendicular to the field can be consistently described by a random walk with the same scattering mechanism as for the propagation along the field.
Therefore, assuming that preformed pairs exist in a superconductor above ${\rm T_c}$~\cite{yang08}, our data indicate that steady propagation of bound pairs may not be realized as long as paring and dissipation emanate from identical degrees of freedom.

Our results on the decay of the bound state may as well lead to a broader understanding of driven strongly correlated systems.
We found a significant difference in out--of--equilibrium response between quasi--1D and 2D systems.
In this context, we showed that at short times after switching on the field the perpendicular distance between carriers $d_{\perp}(t)$ is universally  determined by the distance travelled by the center of mass of two carriers $x(t)$. 
Due to additional decay channels that open as a consequence of charge motion  along the perpendicular direction, the characteristic decay time on the square lattice is much shorter than on the ladder system.
We expect that such cooperative correlation between parallel and perpendicular dynamics may also manifest itself in various setups driven away from equilibrium (like that in Ref.~\cite{mandt11}) where charge carriers initially form a state with inhomogeneous microscopic structure.%

\begin{acknowledgments}
Authors acknowledge stimulating discussions with P. Prelov\v sek.
J.B. and L.V. acknowledge support by the P1-0044 of ARRS, Slovenia.
J.B expresses gratitude for the  support of CINT user program, Los Alamos National Laboratory, NM USA and Gordon Godfrey bequest of UNSW,  Sydney Australia where part of this work has been performed.
M.M. acknowledges support from the N N202052940 project of NCN.
\end{acknowledgments}

\bibliography{references}

\end{document}